\documentclass[letterpaper]{article} 
\usepackage{aaai24}  
\usepackage{times}  
\usepackage{helvet}  
\usepackage{courier}  
\usepackage[hyphens]{url}  
\usepackage{graphicx} 
\usepackage{natbib}  
\usepackage{caption} 
\usepackage{algorithm}
\usepackage{algorithmic}
\usepackage{amsmath}
\usepackage{multirow}
\usepackage{multicol}

\usepackage{amssymb}
\usepackage{amsthm}

\newtheorem{theorem}{Theorem}[section]

\newcounter{subassumption}[assumption]

\makeatletter
\renewcommand{\p@subassumption}{\theassumption}
\makeatother

\usepackage{newfloat}
\usepackage{listings}
\DeclareCaptionStyle{ruled}{labelfont=normalfont,labelsep=colon,strut=off} 
\floatstyle{ruled}
\newfloat{listing}{tb}{lst}{}
\floatname{listing}{Listing}
\title{UNEX-RL: Reinforcing Long-Term Rewards in Multi-Stage Recommender Systems with UNidirectional EXecution}
\author {
    Gengrui Zhang\textsuperscript{\rm 1}\equalcontrib,
    Yao Wang\textsuperscript{\rm 1}\equalcontrib,
    Xiaoshuang Chen\textsuperscript{\rm 1}\equalcontrib,
    Hongyi Qian\textsuperscript{\rm 2},
    Kaiqiao Zhan\textsuperscript{\rm 1},
    Ben Wang\textsuperscript{\rm 1}\footnote{Corresponding author.}
}
\affiliations {
    \textsuperscript{\rm 1}Kuaishou Technology, Beijing, China\\
    \textsuperscript{\rm 2}Beihang University, Beijing, China\\
    \{zhanggengrui, wangyiyan, chenxiaoshuang,
    zhankaiqiao, wangben\}@kuaishou.com,
    qianhongyi@buaa.edu.cn
}

\usepackage{bibentry}
\begin{document}

\maketitle

\begin{abstract}
In recent years, there has been a growing interest in utilizing reinforcement learning (RL) to optimize long-term rewards in recommender systems. Since industrial recommender systems are typically designed as multi-stage systems, RL methods with a single agent face challenges when optimizing multiple stages simultaneously. The reason is that different stages have different observation spaces, and thus cannot be modeled by a single agent. To address this issue, we propose a novel UNidirectional-EXecution-based multi-agent Reinforcement Learning (UNEX-RL) framework to reinforce the long-term rewards in multi-stage recommender systems.
We show that the unidirectional execution is a key feature of multi-stage recommender systems, bringing new challenges to the applications of multi-agent reinforcement learning (MARL), namely the observation dependency and the cascading effect. To tackle these challenges, we provide a cascading information chain (CIC) method to separate the independent observations from action-dependent observations and use CIC to train UNEX-RL effectively. We also discuss practical variance reduction techniques for UNEX-RL. Finally, we show the effectiveness of UNEX-RL on both public datasets and an online recommender system with over 100 million users. Specifically, UNEX-RL reveals a 0.558\% increase in users' usage time compared with single-agent RL algorithms in online A/B experiments, highlighting the effectiveness of UNEX-RL in industrial recommender systems.
\end{abstract}

\section{Introduction}
\begin{figure}
    \centering
    \includegraphics[width=\columnwidth,trim=0 40 0 0, clip]{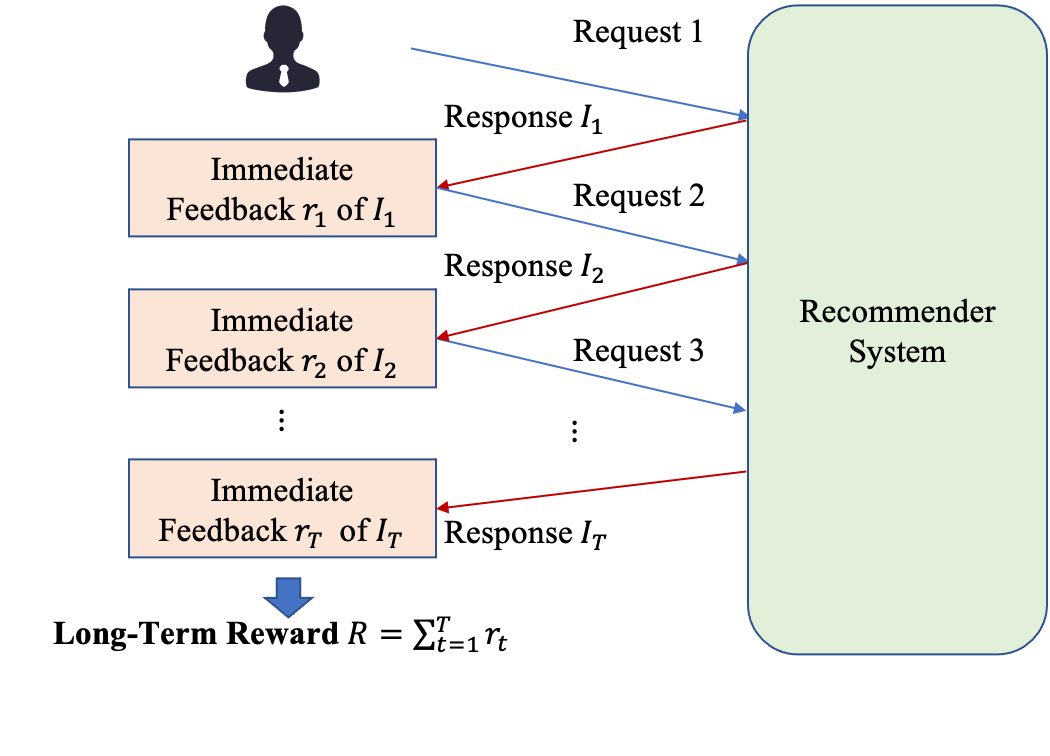}
    \caption{Long-term rewards in  recommender systems.}
    \label{fig:sequential-recommender-systems}
\end{figure}

Recent recommender systems have achieved great success in optimizing immediate engagement, such as click-through rates \cite{pi2020search,lian2018xdeepfm}. However, in real-life applications, it is also necessary to improve the long-term reward, including the total watch time, the session length, the users' retention, etc. The long-term reward is more desirable than immediate engagement because it directly affects some important operational metrics, e.g., daily active users (DAUs) and dwell time \cite{cai2023reinforcing}. Different from the immediate engagement, which has been extensively investigated \cite{wang2021dcn,zhou2018deep,zhou2019deep,pi2020search,lian2018xdeepfm}, it is more challenging to improve the long-term reward in recommender systems, since the long-term reward is affected by users' overall satisfaction of the entire set of recommended items \cite{xue2022resact,wang2022surrogate}, and it is difficult to relate the long-term reward to a single recommended item. Figure \ref{fig:sequential-recommender-systems} shows the long-term reward of a recommender system.

Recently, a growing literature has focused on applying reinforcement learning (RL) to recommender systems due to its ability to improve the long-term reward \cite{chen2019top,gao2022kuairec,zou2019reinforcement,cai2023reinforcing}. RL-based methods treat users as the environment and the recommender system as the agent, and then model the sequential interactions between users and the recommender system as Markov Decision Processes (MDP). After such modeling, several kinds of RL methods \cite{fan2020theoretical,lillicrap2015continuous,fujimoto2018addressing} can be employed to optimize users' long-term rewards.

Despite the growing interest in  RL-based recommendation, it is still challenging to apply RL to industrial-level recommender systems. A significant challenge arises from the fact that industrial recommender systems are typically built upon a multi-stage structure, including matching, pre-ranking, ranking, and re-ranking, to provide real-time recommendations from tens of millions of candidate items in a low latency \cite{wang2020cold}. The upstream stages utilize lightweight models to provide a subset of the candidates to the downstream stages for more precise estimation. Cooperation of the multiple stages is essential for optimizing the long-term rewards \cite{zhang2023rethinking}. However, it is challenging to apply RL to the joint optimization of multiple stages due to the fact that different stages have different observation spaces, e.g., the predictions or statistics of the candidate sets at each stage. This feature makes it inappropriate to model multiple stages in a single-agent RL model.


Recently, multi-agent reinforcement learning (MARL) has reached success in RL problems with multiple agents. Particularly, the centralized training with decentralized execution (CTDE) paradigm has made significant progress in cooperative MARL tasks \cite{rashid2020monotonic}. However, challenges arise when training MARL in multi-stage recommender systems. The key problem is that CTDE assumes that observations of all the agents are simultaneously sampled from the replay buffer in the training stage, but in a multi-stage recommender system, even a slight change of the actions in upstream stages may cause different candidates in downstream stages, leading to different observations and actions in downstream stages. This feature contradicts the assumption of CTDE and deteriorates the performance when training MARL with CTDE. 

This paper proposes a novel UNidirectional-EXecution-based multi-agent Reinforcement Learning (UNEX-RL) framework to apply MARL to multi-stage recommender systems. The unidirectional execution is a unique feature of multi-stage recommender systems compared to typical MARL applications. It leads to two effects in the training process of MARL, namely the observation dependency (OD) in critic learning and the cascading effect (CE) in actor learning. OD and CE deteriorate the performance of traditional CTDE-based approaches in the training of MARL. To mitigate these two effects, we propose a novel cascading information chain (CIC) approach to train UNEX-RL effectively. Moreover, we provide practical variance reduction techniques to further improve the performance. UNEX-RL shows significant improvement in optimizing the long-term rewards in multi-stage recommender systems.


In summary, our contributions are as follows: 
\begin{itemize}
    \item We introduce a novel UNEX-RL framework to improve the long-term rewards of multi-stage recommender systems. To the best of the authors' knowledge, we are the first to investigate the application of RL in industrial multi-stage recommender systems.
    \item We show that the unidirectional execution in multi-stage recommender systems leads to two effects, namely OD and CE. Then we provide a CIC approach to solve these problems. CIC shows a significant improvement over traditional CTDE-based training methods of MARL.
    \item We also discuss practical variance reduction techniques, i.e. stopping gradient (SG) and category-quantile rescaling (CQR), to improve the performance of UNEX-RL.
    \item We conduct extensive experiments in both public datasets and an online recommender system serving over 100 million users. UNEX-RL achieved a 0.558\% increase in users’ usage time in online A/B experiments of a real-world multi-stage recommender system.
\end{itemize}

\section{Related Work}

\subsection{RL in Recommender Systems}
Due to the ability to improve cumulative reward, RL has gathered increasing attention in the realm of recommender systems \cite{shani2005mdp,zou2019reinforcement,zheng2018drn,chen2019top}. RL models multiple interactions between users and recommender systems as MDP, aiming to optimize the long-term cumulative reward of the recommender system. Value-based approaches usually estimate users' preference for each item by Q networks, and select the top k items for users\cite{zheng2018drn, zhao2018recommendations}. Policy-based approaches usually use actor-critic-based methods to learn a policy that aims to improve users' satisfaction\cite{cai2023reinforcing, cai2023two,xue2022resact,xue2023prefrec}. However, previous studies focused on improving a certain part of recommender systems with a single agent, and did not consider the multi-stage structure of industrial-level recommender systems, which is the contribution of our work.

\subsection{MARL}
In fully cooperative MARL, agents are trained to collaborate and achieve better cooperation by sharing the same reward. The CTDE paradigm learns a centralized critic with decentralized actor, has been increasingly used in recent years \cite{foerster2018counterfactual,peng2021facmac,wang2020dop}. In contrast, communication-based methods are used to transfer information between agents to facilitate the completion of cooperative tasks \cite{sukhbaatar2016learning,foerster2016learning}. However, there are still challenges when applying MARL to multi-stage recommender systems (see Sec. \ref{sec:method}), and to the best of the authors' knowledge, we are the first to discuss and solve these challenges.

\section{Preliminary}
\subsection{Multi-Stage Recommender System} \label{sec:multi-stage-recommender-system}
\begin{figure}
    \centering
    \includegraphics[width=\columnwidth, trim=0 5 0 0, clip]{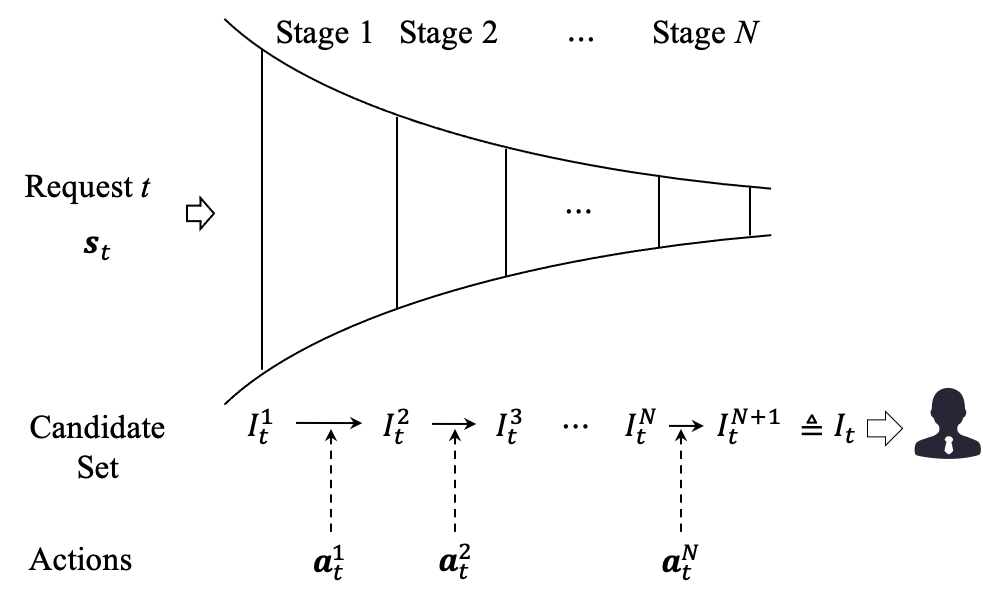}
    \caption{A multi-stage recommender system.}
    \label{fig:multi-stage-abstract}
\end{figure}

A typical multi-stage recommender system includes the matching, pre-ranking, ranking, and re-ranking processes. For the sake of convenience, we utilize the term ``stages" to uniformly describe these processes in the following discussions.
Specifically, there are $N$ stages in the system, as shown in Figure \ref{fig:multi-stage-abstract}. The system receives a user request at each time step $t$, with an observable user state $\boldsymbol{s}_t$, including the user profile, the browsing history, etc. The system returns an item set $I_t$ out of the universal candidate set $I_U$. The $i$-th stage takes a candidate set $I_{t}^i$ as input, and outputs a subset $I_{t}^{i+1}$. We have
$
    I_U = I_t^{1} \supset I_t^2 \cdots \supset I_t^N \supset I_t^{N+1} = I_t
$.

\begin{figure*}[t]
    \centering
    \includegraphics[width=\textwidth, trim=0 2 0 0, clip]{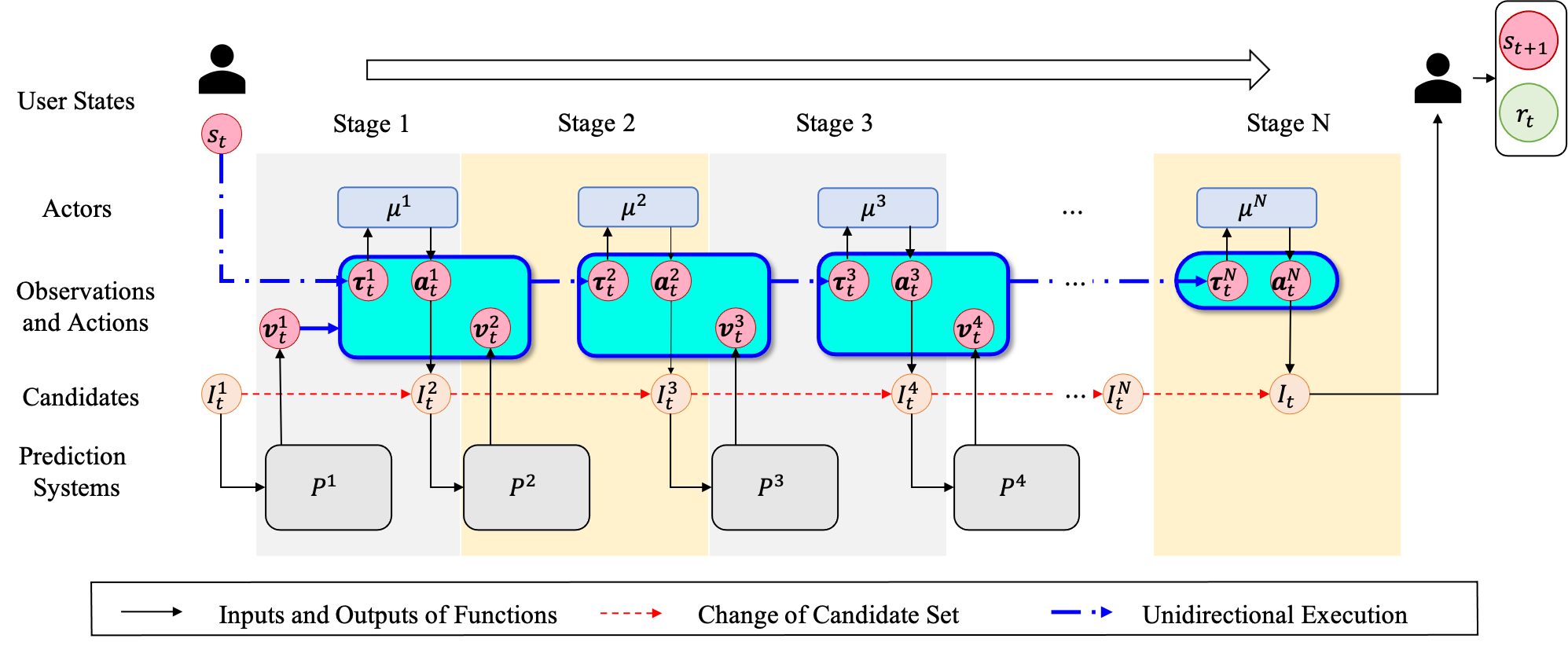}
    \caption{Overall framework of UNEX-RL.}
    \label{fig:UNEX-RL-framework}
\end{figure*}

Generally, Stage $i$ can be modeled as a selecting function $\mathcal{F}^i$ with input $I_t^i$ and output $I_t^{i+1}$, parameterized by $\boldsymbol{a}_t^{i}$:
\begin{equation} \label{eq:n-th-stage}
I^{i+1}_{t} = \mathcal{F}^i(I^{i}_{t}; \boldsymbol{a}^i_{t}),
\end{equation}
where $\boldsymbol{a}_t^i$ is to be determined. We would like to emphasize that Eq. \eqref{eq:n-th-stage} is a very general description of Stage $i$. Here we provide an example:

\noindent\textbf{Example} 1: This example is a simple multi-objective ranking module. The $\mathcal{F}^i$ is the combination of the following step:
\begin{enumerate}
    \item Predicting the user engagement on each item, e.g., the click-through rate, the watch time, etc. The predictions are denoted by $\boldsymbol{p}^i_t$, a $\left|I_{t}^i\right|\times M^i$ matrix, where $M^i$ is the number of predictions.  $\boldsymbol{p}^i_t$ can be obtained from any DNN model.
    \item Taking a linear combination of $\boldsymbol{p}_t^i$ with the weight vector $\boldsymbol{a}_t^i \in R^{M^i}$, i.e. $\boldsymbol{q}^i_t=\boldsymbol{p}_t^i\boldsymbol{a}^i_t \in R^{\left|I_{t}^i\right|}$.
    \item Sorting the items by $\boldsymbol{q}^i_t$ and output the top items as $I_t^{i+1}$.
\end{enumerate}

In this example, the action $\boldsymbol{a}^i_t$ acts as the weights to merge the predictions of each item in $\boldsymbol{p}_t^i$.

\subsubsection{Problem Formulation}
We focus on the optimization of the actions $\boldsymbol{a}_t^i$ at all the stages and all the steps, aiming at improving the user's long-term reward. Formally we have:
\begin{equation} \label{eq:r-t}
    \max_{\boldsymbol{a}_{t}^{1:N}} R_t = \sum_{t'=t}^{\infty}\gamma^{t'-1}r_{t'},
\end{equation}
where the upper bound ``$\infty$'' indicates a summation of the rewards until the user leaves, and $\gamma$ is the discount factor.

MARL-based approaches are usually used to improve the long-term reward with multiple cooperative agents. Before proposing our UNEX-RL framework, we first provide some basic concepts of MARL.

\subsection{MARL}
Agents in cooperative MARL aim to jointly maximize their collective rewards. It can be modeled as a decentralized partially observable Markov decision process (Dec-POMDP) \cite{oliehoek2016concise}, consisting of a tuple $\mathcal{G} = \{ \mathcal{N}, \mathcal{S}, \mathcal{A}, P, R, \Omega, O, i, \gamma\}$, where $\mathcal{N}$ is the set of $N$ agents, $\mathcal{S}$ is the state space, $\mathcal{A}$ is the action space, $P$ is the transition function, $R$ is the reward function,  $\Omega$ is the observation space, and $O$ is the observation function.
At Step $t$, the environment falls into a state $\boldsymbol{s}_t\in\mathcal{S}$, and each agent $i \in \mathcal{N}$ receives an observation $\boldsymbol{\tau}_{t}^i$ drawn according to the observation function $O(\boldsymbol{s}_t, i)$. Each agent learns a policy function $\mu^i$ parameterized by $ \theta^i$, and outputs the action $\boldsymbol{a}_{t}^i \in \mathcal{A}$ as
\begin{equation} \label{eq:policy-function}
    \boldsymbol{a}_{t}^i = \mu^i\left(\boldsymbol{\tau}_{t}^i;\theta^i\right).
\end{equation}
After executing the actions of $N$ agents, the system transits to the next state according to the transition function: $P(\boldsymbol{s}_{t+1}|\boldsymbol{s}_t,\boldsymbol{a}_{t}^{1:N})$ and returns the reward $r_t = R(\boldsymbol{s}_t,\boldsymbol{a}^{1:N}_t)$. Here we use ``$1:N$'' to represent a traverse from 1 to $N$. 


CTDE is a training paradigm of MARL with a centralized critic and decentralized actors, and MADDPG \cite{lowe2017multi} is one of the representative examples. MADDPG learns a centralized critic $Q^i$ for each agent $i$ by sharing the global observation and actions in the training stage:
\begin{equation} \label{eq:global-critic-ctde}
\begin{aligned}
    Q^i(\boldsymbol{\tau}^{1:N}_t, \boldsymbol{a}^{1:N}_t) =\mathbf{E}_{\boldsymbol{\tau}^{1:N}_{t+1:\infty},\boldsymbol{a}^{1:N}_{t+1:\infty}}\left[R^i_t | \boldsymbol{\tau}^{1:N}_t,\boldsymbol{a}^{1:N}_t\right],
\end{aligned}
\end{equation}
where  $R^i_t$
is the long-term reward of Agent $i$ defined in Eq. \eqref{eq:r-t}.
Given $Q^i$, the loss of critic learning in CTDE writes:
\begin{equation}  \label{eq:maddpg-critic}
loss(\phi^i) = \mathbf{E}_D[(y^i-Q^i(\boldsymbol{\tau}^{1:N}_{t}, \boldsymbol{a}^{1:N}_{t}; \phi^i))^2],
\end{equation} 
where
\begin{equation} \label{eq:critic-td}
y^i= r^i_{t} + \gamma*Q^i(\boldsymbol{\tau}^{1:N}_{t+1}, \boldsymbol{a}^{1:N}_{t+1};\phi^{i-}),
\end{equation}
and $\phi^i$ is the parameters of $Q^i$ for Agent $i$. $\boldsymbol{\tau}^{1:N}_{t+1}$ is the observation of Step $t+1$, which is obtained from the replay buffer, and $\boldsymbol{a}^i_{t+1} =\mu^i(\boldsymbol{\tau}^i_{t+1}; \theta^{i-})$ is the action of target policy with delayed parameters $\theta^{i-}$. $\phi^{i-}$ is the parameters of the target critic, and $D$ is the replay buffer. The policy gradient is calculated separately for each agent:
\begin{equation} \label{eq:maddpg-policy-gradient}
\begin{aligned}
\nabla_{\theta^i}J=\mathbf{E}_D\left[\nabla_{\theta^i}\mu^i\nabla_{\mu^i}Q^i\left.\left(\boldsymbol{\tau}^{1:N}_{t},\boldsymbol{a}^{1:N}_{t}\right)\right|_{\boldsymbol{a}^i_t = \mu^i\left(\boldsymbol{\tau}^i_t\right)}\right]
\end{aligned}
\end{equation} 
where $\boldsymbol{a}^i_t$, the action of agent $i$, is obtained of the policy $\mu^i$, while the other actions are sampled from the replay buffer.


\section{Method} \label{sec:method}
This section provides the UNEX-RL method. Firstly, we provide the overall framework and show that the unidirectional execution leads to several challenges i.e. OD and CE. Then, we propose CIC to tackle these challenges. Finally, practical variance reduction techniques are provided to reduce the variance and to improve the performance.

\subsection{Overall Framework}
The UNEX-RL framework is illustrated in Figure \ref{fig:UNEX-RL-framework}. When a user opens the app, a session begins, which consists of multiple requests until the user leaves the app. At Step $t$, the system obtains a user state $\boldsymbol{s}_t$ together with the first candidate set $I_{t}^1$. The $N$ stages are modeled as $N$ agents. Agent $i$ receives the information available at Stage $i$, denoted by $\boldsymbol{\tau}^i_{t}$, and uses the policy function $\mu^i$ to provide the action $\boldsymbol{a}_{t}^i$. The action $\boldsymbol{a}_t^i$ is used to select $I_{t}^{i+1}$ out of $I_{t}^i$ (see Eq. \eqref{eq:n-th-stage}). After the $N$ stages, the system outputs the final recommended set $I_t$. Then, the user provides feedback $r_t$ and determines whether to send the next request or leave. UNEX-RL aims to maximize the long-term reward $R_t$ defined in Eq. \eqref{eq:r-t}.

There are three parts in the  agent observation $\boldsymbol{\tau}^i_t$:
\begin{itemize}
    \item The inherit part $\boldsymbol{\tau}_t^{i-1}$ when $i > 1$, which means Stage $i$ possesses all the observations in Stage $i - 1$.
    \item The action of the previous stage, i.e. $\boldsymbol{a}_t^{i-1}$.
    \item The information obtained in this stage for the first time, e.g., new predictions or statistics about the candidate set $I_t^i$. We denote it as $\boldsymbol{v}_t^i$, and generally, we assume $\boldsymbol{v}_t^i$ to be obtained from an information extraction process:
    \begin{equation} \label{eq:v-t}
        \boldsymbol{v}_t^i = \mathcal{P}^i\left(I_t^i;P^i\right),
    \end{equation}
    where $P^i$ is the parameter. We do not explicitly model $\mathcal{P}^i$, but we assume it to be accessible in training.
\end{itemize}

Formally we have:
\begin{align}
\boldsymbol{\tau}_{t}^1 = \left[\boldsymbol{s}_t,\boldsymbol{v}_t^1\right], \boldsymbol{\tau}_{t}^{i+1} =\left[\boldsymbol{\tau}_{t}^i,\boldsymbol{a}_{t}^i, \boldsymbol{v}_t^{i+1}\right].\label{eq:UNEX-RL-deduction}
\end{align}

The \textbf{unidirectional execution}, which is the key difference between UNEX-RL and traditional MARL, arises from the information extraction process $\mathcal{P}^i$ in Eq. \eqref{eq:v-t}. Since $\boldsymbol{v}^i_t$ must be obtained after the candidate set $I_t^i$ is determined, Agent $i$ must be executed after Agent $i - 1$, and all the agents must be executed in series (see  the blue arrows in Figure \ref{fig:UNEX-RL-framework}).

MARL is usually trained by CTDE \cite{lowe2017multi}, but the unidirectional execution brings two challenges:
\begin{itemize}
    \item \textbf{Observation Dependency (OD)}: The OD problem is about the critic learning, i.e., Eq. \eqref{eq:maddpg-critic}\eqref{eq:critic-td}. In the calculation of target actions in Eq. \eqref{eq:critic-td}, the observations $\boldsymbol{\tau}^{1:N}_{t+1}$ are sampled from the replay buffer, and the actions $\boldsymbol{a}^{1:N}_{t+1}$ are calculated via the target policy. However, if we change the actions $\boldsymbol{a}^i_{t+1}$, the observations $\boldsymbol{\tau}^{i+1}_{t+1}$ in Stage $i+1$ will also change according to Eq. \eqref{eq:UNEX-RL-deduction}, which contradicts the assumption that $\boldsymbol{\tau}^{1:N}_{t+1}$ is predefined in the replay buffer.
    \item \textbf{Cascading Effect (CE)}: The CE problem is about the actor learning part, i.e., Eq. \eqref{eq:maddpg-policy-gradient}, which optimizes the policy function $\mu^i$ of the $i$-th agent under the condition that the actions of other agents are sampled from the replay buffer. However, according to Eq. \eqref{eq:UNEX-RL-deduction}, the action $\boldsymbol{a}^i_{t}$ is contained in the observation of downstream stages, i.e., $\boldsymbol{\tau}^{i+1:N}_t$, hence will affect the decisions of downstream stage actors, which is not taken into account in Eq. \eqref{eq:maddpg-policy-gradient}.
\end{itemize}



\subsection{Training of UNEX-RL}
This subsection presents the training of UNEX-RL, especially the solution to the OD and CE problems, as shown in Algorithm \ref{alg:training}. We use the actor-critic structure. Different from Eq. \eqref{eq:maddpg-critic}\eqref{eq:critic-td}, all the agents share the reward $r_t$ from the user, and thus we only use a global critic $Q^g$.

\begin{figure}
    \centering
    \includegraphics[width=\columnwidth, trim=0 5 0 0, clip]{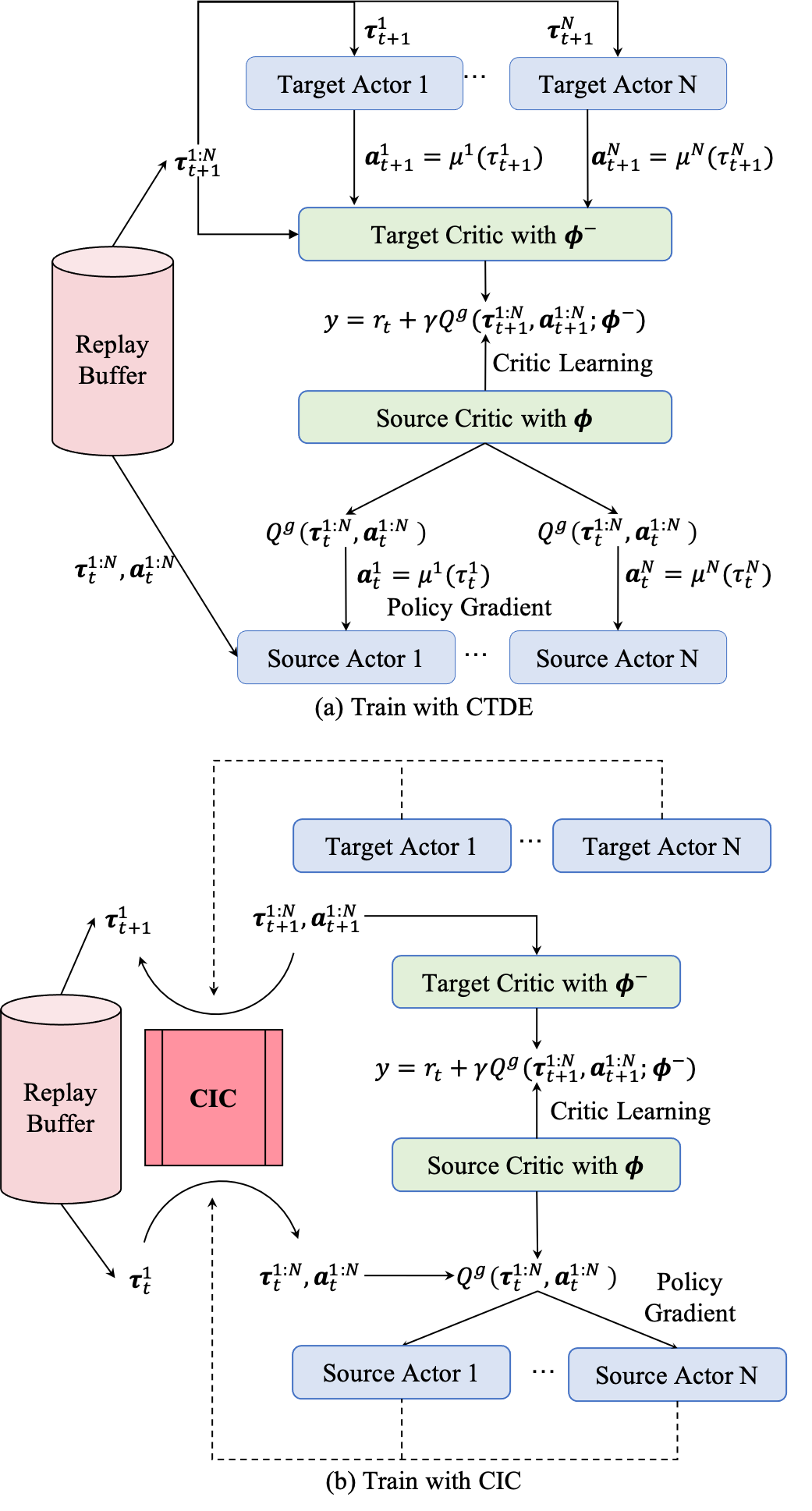}
    \caption{Train with CTDE and CIC.}
    \label{fig:unidirectional-cooperation}
\end{figure}

\begin{algorithm*}
\caption{The training process of UNEX-RL.}
\label{alg:training}
\begin{algorithmic}[1]
\STATE Input: $\left\{\boldsymbol{s}_{1:T}, \boldsymbol{\tau}^{1:N}_{1:T}, \boldsymbol{a}^{1:N}_{1:T}, r_{1:T}\right\}$ for each user.
\STATE Output: A critic function $Q^g\left(\boldsymbol{\tau}^{1:N}_t, \boldsymbol{a}^{1:N}_t;\phi\right)$ parameterized by $\phi$; $N$ policies $\mu_i\left(\boldsymbol{\tau}^i_t;\theta^i\right)$, parameterized by $\theta^i, 1\leq i\leq N$.
\FOR{each user session with $T$ requests from the replay buffer}
    \FOR{$t = 1,\cdots, T$}
        \STATE Collect the reward $r_t$, the observation $\boldsymbol{\tau}^{1:N}_t$, and all the actions $\boldsymbol{a}^{1:N}_t$ from the replay buffer.
        \STATE Collect the first-stage observation of the next step $\boldsymbol{\tau}^1_{t+1}$ from the replay buffer.
        \STATE Calculate the CIC  according to Eq. \eqref{eq:cic}  to obtain the target action $\boldsymbol{a}^{1:N}_{t+1}$ and the updated observation $\boldsymbol{\tau}^{1:N}_{t+1}$ from $\boldsymbol{\tau}_t^1$.
        \STATE Critic learning: $\phi \leftarrow \phi - \alpha \nabla_{\phi}\mathcal{L}$, where $\mathcal{L}(\phi)$ is defined in Eq. \eqref{eq:UNEX-RL-critic}\eqref{eq:UNEX-RL-y}, and $\alpha$ is the learning rate.
        \STATE Actor learning: $\theta^i \leftarrow \theta^i - \beta \nabla_{\theta^i}J_i$, where $\nabla_{\theta^i}J_i$ is defined in \eqref{eq:UNEX-RL-policy-gradient}, and $\beta$ is the learning rate.
    \ENDFOR
\ENDFOR
\end{algorithmic}
\end{algorithm*}


We first consider the solution to OD. A key dilemma is that the observations from the replay buffer contradict the observations of target actors in the training stage. Thus, we need to find an independent subset of the observations. Our solution is based on the following finding: given all the target actors and all the checkpoints of the system, we are able to replay the total recommendation only based on the first observation in the training stage. Formally we have:
\begin{theorem} \label{thm:closure}
Denote $\mathcal{E}(\cdot)$ as the information implied by input variables. Assume that the parameters $P^i$ of the information extraction in Eq. \eqref{eq:v-t} are given for all $i$. Then $\forall i, 2\leq i \leq N$, the set $\left\{\boldsymbol{\tau}^{1}_t, \boldsymbol{a}^{1:i-1}_t\right\}$ contains all the information of the observation $\boldsymbol{\tau}^{i}_t$, i.e. $\mathcal{E}\left(\boldsymbol{\tau}^{i}_t\right)\subset \mathcal{E}\left(\left\{\boldsymbol{\tau}^1_t, \boldsymbol{a}^{1:i-1}_t\right\}\right)$. 
\end{theorem}

Please refer to Appendix. \ref{appendix:proof-closure} for the detailed proof of Theorem \ref{thm:closure}.
Note that the assumption of a given parameter $P^i$ is reasonable since it is common to replay the online recommendation processes based on model checkpoints in the offline training \cite{chen2019top}.

Theorem \ref{thm:closure} models the information flow of the unidirectional execution in Fig. \ref{fig:UNEX-RL-framework}. Based on this finding, we propose a cascade information chain (CIC) method, described in Algorithm \ref{alg:cic}, to obtain $\boldsymbol{\tau}_t^{1:N}$ and $\boldsymbol{a}_t^{1:N}$ only from the first observation $\boldsymbol{\tau}_t^1$. CIC executes iteratively, leveraging the information of upstream stages to obtain the observations of downstream stages until the whole observations and actions have been obtained.
For simplicity, we represent CIC as:
\begin{equation}\label{eq:cic}
\left[\boldsymbol{\tau}_t^{1:N},\boldsymbol{a}_t^{1:N}\right] = C\left(\boldsymbol{\tau}_t^1;\theta^{1:N},P^{1:N}\right).
\end{equation}

\begin{algorithm}[t]
\caption{The process of CIC.}
\label{alg:cic}
\begin{algorithmic}[1]
\STATE Input: The first observation $\boldsymbol{\tau}_t^1$, actor parameters $\theta^{1:N}$, and prediction system parameters $P^{1:N}$.
\STATE Output: Actions $\boldsymbol{a}_t^{1:N}$ and observations $\boldsymbol{\tau}_t^{1:N}$.
\STATE $\boldsymbol{a}_t^1 = \mu^1\left(\boldsymbol{\tau}_t^1;\theta^1\right)$
\FOR{$i=2,\cdots,N$}
    \STATE $I_t^i = \mathcal{F}^i\left(I_t^{i-1};\boldsymbol{a}_t^{i-1}\right)$, as shown in Eq. \eqref{eq:n-th-stage}.
    \STATE $\boldsymbol{v}_t^i = \mathcal{P}^i\left(I_t^i;P^i\right)$, as shown in Eq. \eqref{eq:v-t}.
    \STATE $\boldsymbol{\tau}_t^i = \left[\boldsymbol{\tau}_t^{i-1},\boldsymbol{a}_t^{i-1},\boldsymbol{v}_t^i\right]$, as shown in Eq. \eqref{eq:UNEX-RL-deduction}.
    \STATE $\boldsymbol{a}_t^i = \mu^i\left(\boldsymbol{\tau}_t^i;\theta^i\right)$
\ENDFOR
\end{algorithmic}
\end{algorithm}


Via CIC, the critic learning can be formulated as:
\\
\begin{equation} \label{eq:UNEX-RL-critic}
\mathcal{L}(\phi) = \mathbf{E}_D[(y^{g}-Q^{g}(\boldsymbol{\tau}^{1:N}_t, \boldsymbol{a}^{1:N}_t; \phi))^2],
\end{equation} 
where
\begin{equation} \label{eq:UNEX-RL-y}
y^{g}_t = r_t + \gamma*Q^{g}(C\left(\boldsymbol{\tau}_{t+1}^1;\theta^{1:N-},P^{1:N}\right);\phi^{-}),
\end{equation}
$\phi$ and $\phi^{-}$ are parameters of the critic and the target critic. Here the only independent observation which needs to be obtained from the replay buffer is $\boldsymbol{\tau}^1_{t+1}$, while the observations of the downstream stages are obtained from CIC. Therefore, CIC provides a solution to the OD problem.

Now we discuss the actor learning of UNEX-RL and show that CIC can also be used to solve the CE problem. Note that traditional CTDE in Eq. \eqref{eq:maddpg-policy-gradient} does not consider the influence of $\boldsymbol{a}_t^i$ to the actions of its downstream agents. In contrast, any change of $\mu^i$ will change not only the corresponding action $\boldsymbol{a}_t^i$, but also the actions of succeeding stages $\boldsymbol{a}_t^{i+1:N}$ in CIC. Therefore, we just need to use CIC in the policy gradient:
\begin{equation} \label{eq:UNEX-RL-policy-gradient}
    \nabla_{\theta^i}J_i = \mathbf{E}_D\left[\nabla_{\theta^i}\mu^i\nabla_{\mu^i}Q^g\left(C\left(\boldsymbol{\tau}_t^1;\theta^{1:N},P^{1:N}\right)\right)\right]
\end{equation}
 
In a word, we use CIC to replay the execution process in the training stage to solve OD and CE problems. The difference between CTDE and CIC  is shown in Figure \ref{fig:unidirectional-cooperation}.
\subsection{Variance Reduction Techniques}
RL in recommender systems suffers from large variance due to the highly sparse data \cite{chen2019top}, and MARL makes things even worse due to the decentralized actions of multiple agents \cite{lowe2017multi}. Here we provide some practical variance reduction techniques.
\subsubsection{Stopping Gradient (SG)}
The CIC function provides multiple paths from $Q^g$ to $\mu^i$, which will make the gradient somehow uncontrollable in backward propagation. Here we stop the gradient from 
$\mu^{i+1:N}$ to $\mu^i$, and only use the direct gradient from $Q^g$ to $\mu^i$ in critic learning.
\subsubsection{Category Quantile Rescale (CQR)}
The user's feedback $r_t$ depends not only on the user's interest in the recommended item but also on the user's or the item's bias feature. Such bias leads to a large range of $r_t$, resulting in a large variance \cite{zhan2022deconfounding}. Please refer to Appendix. \ref{appendix:bias} for more details about these biases.
Here we propose CQR, a reward-reshaping method to solve this problem. We obtain the group of users similar to the user $u$, denoted by $\mathcal{G}_u$, and the group of items similar to the item $i$, denoted by $\mathcal{G}_i$. There are sufficient discussions on how to find similar users and items \cite{pi2020search,chen2021end}. The CQR reward is the quantile of the reward conditioned on $\mathcal{G}_u$ and $\mathcal{G}_i$:
\begin{equation} \label{eq:cqr}
\tilde{r}_{t,ui} = \Phi\left(r_{t,ui} | \mathcal{G}_u,\mathcal{G}_i\right)
\end{equation}
where $\Phi$ denotes the cumulative density function.

Clearly, $\tilde{r}_{t,ui}$ obeys a uniform distribution, and this uniformity remains when conditioned on any $\mathcal{G}_u$ and $\mathcal{G}_i$. Thus, the range of $\tilde{r}_{t,ui}$ will be moderate compared with $r_t$, which can reduce the variance of critic learning.

\section{Evaluation}
We conduct extensive offline and online experiments, and answer the following research questions (RQs):
\begin{itemize}
\item \textbf{RQ1}: How does the UNEX-RL framework perform compared to other state-of-the-art (SOTA) methods?
\item \textbf{RQ2}: Does UNEX-RL with CIC effectively solve the problems in multi-stage recommender systems?
\item \textbf{RQ3}: How do the variance reduction techniques contribute to the performance?
\item \textbf{RQ4}: Can the UNEX-RL framework lead to improvements in real-world recommender systems?
\end{itemize}

\subsection{Preparing the Dataset}
\subsubsection{Dataset and Simulator}
We select KuaiRand \cite{gao2022KuaiRand} as the offline experimental dataset. KuaiRand is a public dataset obtained from Kuaishou, containing 27,285 users and 32,038,725 items. It encompasses contextual features of both users and items, along with multiple feedback signals from users. 
\\To emulate real user behavior, we construct a user simulator to act as the environment and mimic the user's interaction with the recommender system. After receiving the recommended items, the user simulator provides immediate feedback and determines whether to send the next request. We follow the work in \cite{xue2023prefrec} to introduce a quitting mechanism: the user will exit the session once they have exhausted their satisfaction, and an early quit will also lead to inferior performance in terms of reward metrics.
\subsubsection{Offline Experiment Details}
Align with industrial recommender systems, we consider three stages i.e., matching, pre-ranking, and ranking. The operation in each stage is the same as Example 1 in Sec. \ref{sec:multi-stage-recommender-system}, i.e., a linear ranking function $\boldsymbol{q}^i_t=\boldsymbol{p}_t^i\boldsymbol{a}^i_t$, where $ \boldsymbol{p}_t^i\in\left|I_{t}^i\right|\times 3$ represents the prediction of three feedbacks, i.e., like, long view and watch time. The action $\boldsymbol{a}_t^i \in R^{3}$ is the weight of the linear ranking. The state $\boldsymbol{s}_t$  comprises the user profile, the behavior history, and the request context. The observation $\boldsymbol{v}_t^i$ includes statistics of the candidate items.
To ensure fairness, we use a consistent network architecture, i.e., a multi-layer perceptron with 5 layers, for actors and critics of all the compared methods. The hyperparameters can be found in Table \ref{tab:hyper-params} in Appendix. \ref{appendix:hyper-params}.

\subsection{Compared Methods}
\begin{itemize}
\item \textbf{Cross Entropy Method (CEM)} \cite{rubinstein2004cross}: a black-box optimization method commonly used for hyper-parameter optimization. We use CEM to search the best parameters $\boldsymbol{a}_t^i$ at each stage.
\item \textbf{DDPG} \cite{lillicrap2015continuous}: Since DDPG is a single-agent method, We deploy it at the final ranking stage.
\item \textbf{TD3} \cite{fujimoto2018addressing}: Similar to DDPG, we deploy TD3 at the final ranking stage.
\item \textbf{UNEX-RL-CTDE}: UNEX-RL trained by CTDE along with CQR in a multi-stage recommender system.
\item \textbf{UNEX-RL-CTDE (w/o CQR)}: UNEX-RL-CTDE without CQR.
\item \textbf{UNEX-RL-CIC}: UNEX-RL trained under CIC in Algorithm \ref{alg:training}, along with SG and CQR.
\item \textbf{UNEX-RL-CIC (w/o CQR)}: UNEX-RL-CIC without CQR.
\item \textbf{UNEX-RL-CIC (w/o SG)}:  UNEX-RL-CIC without SG.

\end{itemize}

\subsection{Evaluation Metric}
We measure two kinds of long-term rewards, namely the \textbf{WatchTime}, i.e., the accumulated watching time of all the items watched in the session, and the \textbf{Session Length}, i.e.,
the number of items watched in the session.

\subsection{Performance Comparison (RQ1)}
The results of the offline  experiments are shown in Table \ref{table:offline exp results}. CEM searches a global weight vector $\boldsymbol{a^i}$ for each stage. Since DDPG and TD3 are single-agent methods, we deploy them at the final ranking stage. Compared to CEM, RL-based algorithms perform much better due to their advantage of optimizing the long-term rewards of the users. Furthermore, UNEX-RL achieves significantly better performance, showing that MARL can better release the ability of RL in a multi-stage recommendation system. Moreover, the comparison between UNEX-RL-CIC and UNEX-RL-CTDE shows the effectiveness of our proposed CIC approach.

\begin{table}[t]
\centering
\begin{tabular}{c|c|c}
    \hline
    \multirow{2}{*}{Methods} & WatchTime & Session \\ 
    & (s) & Length \\
    \hline
    CEM &  654.0 & 15.3 \\
    DDPG & 732.6 & 18.2 \\
    TD3 & 763.2 & 18.9 \\
    \hline
    UNEX-RL-CTDE(w/o CQR) & 782.5 & 19.7 \\
    UNEX-RL-CTDE & 887.2 & 21.6 \\
    \hline
    UNEX-RL-CIC(w/o CQR) & 974.3 & 22.7 \\
    UNEX-RL-CIC(w/o SG) & 752.1 & 18.4  \\
    \textbf{UNEX-RL-CIC} & \textbf{1056.2} & \textbf{24.2} \\\hline
\end{tabular}
\caption{The overall performance of different methods.}
\label{table:offline exp results}
\end{table}

\subsection{Impacts of Multiple Stages (RQ2)}
\begin{figure}
    \centering
    \includegraphics[width=0.38\textwidth]{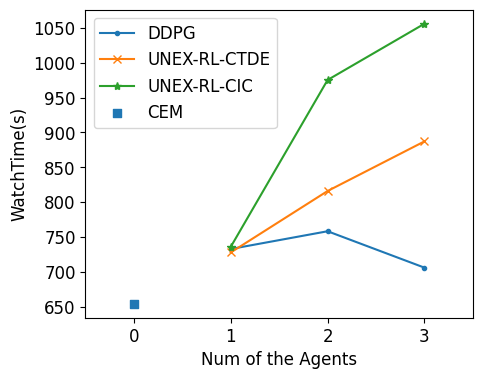}
    \caption{Performance of different numbers of agents.}
    \label{fig:jizhua}
\end{figure}

To demonstrate that the UNEX-RL framework can better adapt to multi-stage recommender systems, we analyze the impact of the number of agents on different methods, as shown in Figure \ref{fig:jizhua}. CEM can be regarded as a method with 0 agents. For all other RL-based methods, we adjust the number of agents from 1 to 3. Specifically for DDPG, we deploy an independent DDPG at each stage. 
\\
When there is only one agent, UNEX-RL-CTDE and UNEX-RL-CIC degrade to DDPG, and the three methods exhibit the same performance. As the number of agents increases, both UNEX-RL-CTDE and UNEX-RL-CIC show significant improvements, proving the effectiveness of UNEX-RL. However, multiple independent DDPGs show even lower performance since the lack of cooperation among the agents. When there are multiple agents, UNEX-RL-CIC outperforms UNEX-RL-CTDE, showing that the CIC training paradigm can better facilitate cooperation among the agents in multi-stage recommender systems.

\subsection{Impacts of Variance Reduction (RQ3)}
\begin{figure}
    \centering
    \includegraphics[width=0.38\textwidth]{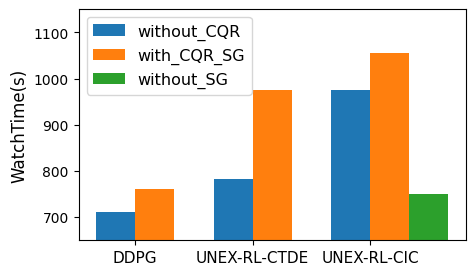}
    \caption{Performance of Variance Reduction Techniques.}
    \label{fig:VR}
\end{figure}

To analyze the contribution of the variance reduction techniques, we deploy the ablation study of UNEX-RL, shown in Figure \ref{fig:VR}. By removing CQR from UNEX-RL-CTDE and UNEX-RL-CIC, the performance significantly deteriorates, proving that CQR can significantly improve the performance of UNEX-RL. Also, by removing SG from UNEX-RL-CIC, the performance will be much worse, showing that SG is a key technique to improve the training of CIC. We do not discuss the impacts of SG on CTDE and DDPG since these two methods do not contain gradients from $\mu^{i+1}$ to $\mu^i$.

\subsection{Online A/B Experiments (RQ4)}
We deploy UNEX-RL on Kwai, a short video app with over 100 million users. Through online experiments, we compare UNEX-RL with the current SOTA methods. The baseline is CEM, and we successively experiment with DDPG, TD3, UNEX-RL-CTDE, and UNEX-RL-CIC. The evaluation metrics are the relative improvements of daily WatchTime and session-wise WatchTime.
\\
In the online experiment, we deploy agents in the pre-ranking and ranking stages, and the two agents return actions as the weight vector of the ranking functions. The observations include user information that all the agents share, and stage-wise observations, i.e., the statistics of the predicted scores in the current stage.
\\
Table \ref{table:live exp results} shows the comparisons of different methods. According to Table \ref{table:live exp results}, DDPG and TD3 show significant improvements  compared to CEM, showing the effectiveness of RL to improve users' long-term rewards. More importantly, both UNEX-RL-CTDE and UNEX-RL-CIC  perform better than the RL methods with a single agent, indicating that the proposed UNEX-RL framework leads to improvements
in real-world recommender systems. Notably, UNEX-RL-CIC achieves the best performance compared to other methods, i.e., a 0.953\% gain of daily WatchTime compared with CEM, and a 0.558\% gain compared with TD3. We emphasize that a 0.1\% improvement  holds statistical significance in our system.

Furthermore, we conduct a 150-day online experiment, with the baseline group remaining with the CEM approach, while the experimental group successively introduced TD3, UNEX-RL-CTDE, and UNEX-RL-CIC. DDPG is not introduced to the experimental group since it is tested in the same period as TD3 with a lower performance. We mark the deployment time of these methods, showing significant gains when deploying UNEX-RL. The results shows the improvement of UNEX-RL over existing methods.
 
 \begin{figure}
    \centering
    \includegraphics[width=0.4\textwidth]{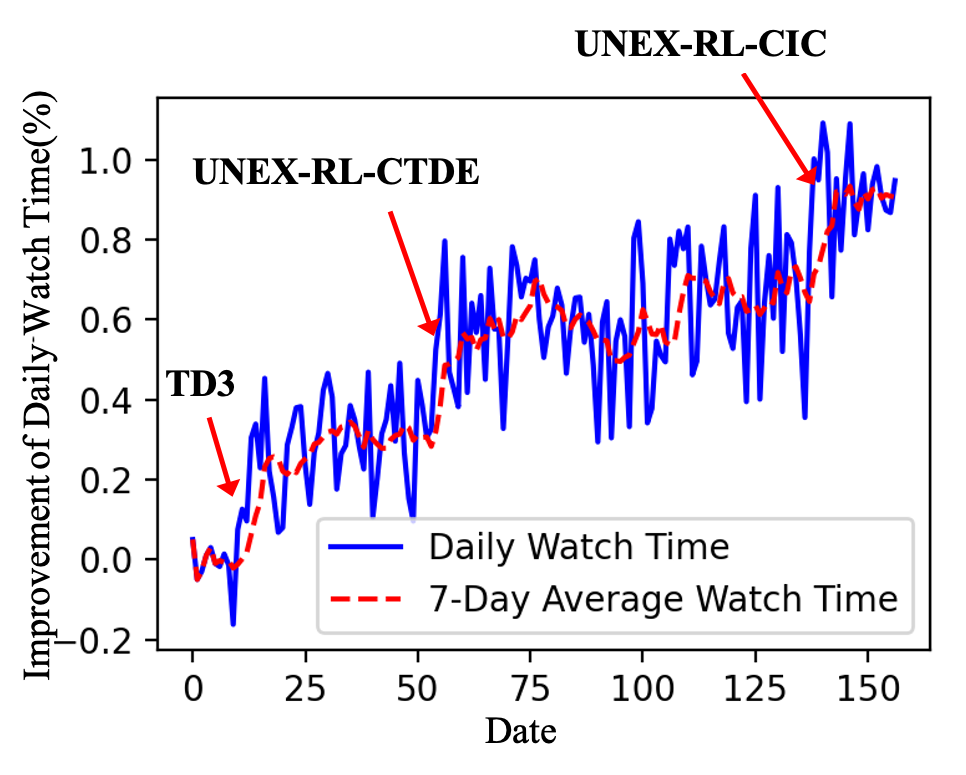}
    \caption{Results of a long period online experiment.}
    \label{fig:ab}
\end{figure}

\begin{table}[t]
\centering
\begin{tabular}{c|c|c}
    \hline
    \multirow{2}{*}{Algorithms} & Session& Daily \\
    &WatchTime & WatchTime\\
    \hline
    CEM &  - &  -  \\
    DDPG & + 0.233\% & + 0.219\%  \\
    TD3 & + 0.414\% & + 0.395\%  \\
    \hline
    UNEX-RL-CTDE & + 0.610\% & + 0.602\%  \\
    \textbf{UNEX-RL-CIC} & \textbf{+ 0.970\%} & \textbf{+ 0.953\%} \\ \hline
\end{tabular}
\caption{Results of online experiments compared with CEM.}
\label{table:live exp results}
\end{table}

\section{Conclusion}
This paper provides a general framework, i.e. UNEX-RL, for applying MARL to multi-stage recommender systems. UNEX-RL differs from traditional MARL in the unidirectional execution, which brings challenges to the training of UNEX-RL. To tackle the challenges of OD and CE arising from the unidirectional execution, we provide the CIC approach to effectively train UNEX-RL, and provide practical variance reduction techniques, i.e. SG and CQR, to further improve the performance. UNEX-RL has shown its effectiveness on both public datasets and a real-world recommender system, and has been deployed online, serving over 100 million users.

\bibliography{AAAI_RL/aaal_rl_camera_ready}

\onecolumn
\newpage

\appendix
\section{Proof of Theorem \ref{thm:closure}} \label{appendix:proof-closure}
\begin{proof}
We extend the theorem to the case that $i=1$, where we set $\left\{\boldsymbol{a}_t^{1:0}\right\}=\varnothing$. When $i=1$, it is trivial that $\boldsymbol{\tau}_t^1\in\mathcal{E}(\boldsymbol{\tau}_t^1)$. Then we prove the theorem by induction.

Suppose the statement holds for $i = k$, i.e. $\boldsymbol{\tau}^{k}_t \in \mathcal{E}\left(\left\{\boldsymbol{\tau}^1_t, \boldsymbol{a}^{1:k-}_t\right\}\right)$, then according to Eq. \eqref{eq:UNEX-RL-deduction}, we have
\begin{equation} \label{eq:thm-induction}
\boldsymbol{\tau}^{k+1}_{t} \in \mathcal{E}\left(\left\{\boldsymbol{\tau}^{k}_t, \boldsymbol{a}^k_t, \boldsymbol{v}^{k+1}_t\right\}\right).
\end{equation}

According to Eq. \eqref{eq:v-t}, $\boldsymbol{v}_t^{k+1}\in\mathcal{E}\left(I_t^{k+1}\right)$. Then, according to Eq. \eqref{eq:n-th-stage}, we have
\begin{equation}
    I_t^{k+1} \in \mathcal{E}\left(\left\{I_t^k,\boldsymbol{a}_t^k\right\}\right) \subset \mathcal{E}\left(\left\{I_t^{k-1},\boldsymbol{a}_t^{k-1:k}\right\}\right) \cdots \subset \mathcal{E}\left(\left\{I_t^1, \boldsymbol{a}_t^{1:k}\right\}\right).
\end{equation}

Since $I_t^1$ is the universal candidate set $I_U$ independent of this request, we have $I_t^{k+1}\in\mathcal{E}\left(\left\{\boldsymbol{a}_t^{1:k}\right\}\right)$. Therefore we have $\boldsymbol{v}_t^{k+1}\in\mathcal{E}\left(\left\{\boldsymbol{a}_t^{1:k}\right\}\right)$.

Then, according to Eq. \eqref{eq:thm-induction}, we have 
\begin{equation} \label{eq:information-k-plus-1}
\boldsymbol{\tau}_t^{k+1}\in\mathcal{E}\left(\left\{\boldsymbol{\tau}_t^k, \boldsymbol{a}_t^{1:k}\right\}\right).
\end{equation}
Finally, by adding $\boldsymbol{\tau}_t^k\in\mathcal{E}\left(\left\{\boldsymbol{\tau}_t^1,\boldsymbol{a}_t^{1:k-1}\right\}\right)$ to Eq. \eqref{eq:information-k-plus-1}, we complete the proof.
\end{proof}

\section{More Details about User and Item Bias} \label{appendix:bias}
\begin{figure}[h]
    \centering
    \includegraphics[width=1.0\textwidth]{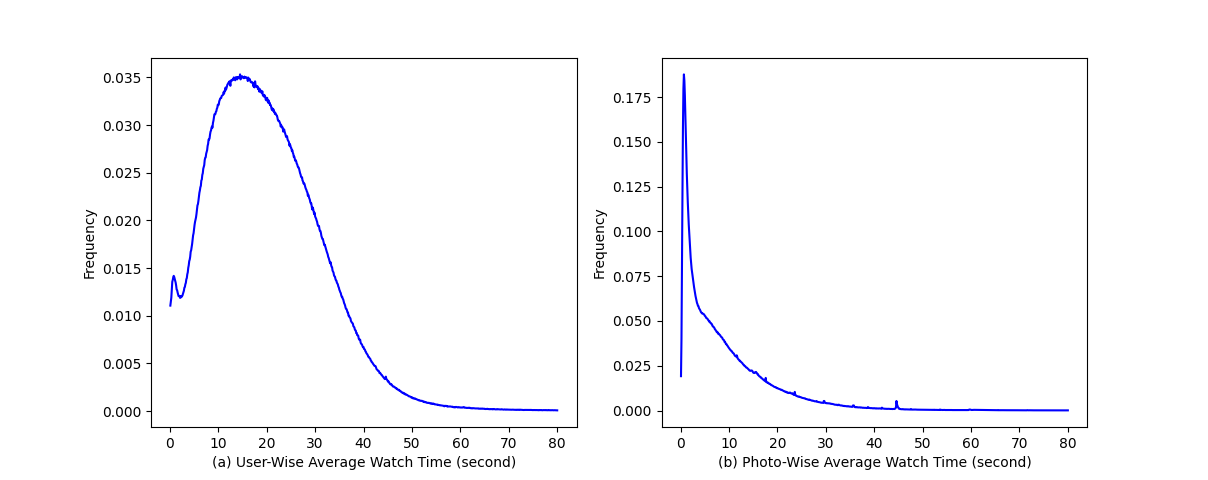}
    \caption{Distribution of user-wise and photo-wise average watch time.}
    \label{fig:bias}
\end{figure}

In RL-based recommender systems, the reward term $r_t$ usually comes from user feedback on the items. However, this feedback depends not only on the users' interest in the recommended items, but also on the users' and items' biases. Here we provide an example of the watch time in an online video recommender system. Figure \ref{fig:bias} shows the distribution of the user-wise average watch time and photo-wise average watch time in the online video recommender system, showing a large range of watch time over different users and different photos. This distribution shows that different users will be used to spending different time on the recommended videos. A similar phenomenon exists in the case of photos. These biases make the reward term span a large range of values.

\newpage
\section{Hyper Parameters} \label{appendix:hyper-params}
\begin{table}[h]
\centering
\begin{tabular}{l|l}
    \hline
    Hyper-parameter & Value \\ \hline
    Matching Model (stage 1) & FM \cite{rendle2010factorization} \\
    Pre-Ranking Model (stage 2) & DSSM \cite{huang2013learning} \\
    Ranking Model (stage 2) & DeepFM \cite{guo2017deepfm} \\
    Number Of Items In Each Stage &[10k, 2k, 1k]\\
    Optimizer & Adam \\
    Actor Learning Rate &  0.0001 \\
    Critic Learning Rate & 0.0002\\
    Number of Agents & 3 \\
    Action Dimensions Of Each Actor & [3, 3, 3] \\
    Discount Factor & 0.9 \\
    Action Upper Bound & 2.0 \\
    Replay Buffer Size & $1*10^{6}$ \\
    Train Batch Size & 1024 \\
    Fine-Tuning & True \\
    Normalized Observations & True\\
    Training Platform    &Tensorflow\\ \hline
\end{tabular}
\caption{The Hyper-parameters of UNEX-RL.}
\label{tab:hyper-params}
\end{table}

\end{document}